\documentclass[english,twocolumn,aps,prl,showpacs,10pt]{revtex4-1}

\usepackage[T1]{fontenc}
\usepackage[latin1]{inputenc}
\usepackage{graphicx}
\usepackage{amssymb}

\usepackage{babel}
\makeatother

\begin{document}

\title{Quantum anomaly and 2D-3D crossover in strongly interacting Fermi gases}

\author{T. Peppler$^1$, P. Dyke$^{1,2}$, M. Zamorano$^1$, S. Hoinka$^{1,2}$, and C. J. Vale$^{1,2,*}$}
\affiliation{$^1$Centre for Quantum and Optical Sciences, Swinburne University of Technology, Melbourne 3122, Australia. \\ $^2$ARC Centre of Excellence in Future Low-Energy Electronics Technologies.\\
{$^\ast$To whom correspondence should be addressed; E-mail: cvale@swin.edu.au} }

\date{\today}
\begin{abstract}
We present an experimental investigation of collective oscillations in harmonically trapped Fermi gases through the crossover from two to three dimensions. Specifically, we measure the frequency of the radial monopole or breathing mode as a function of dimensionality in Fermi gases with tunable interactions. The frequency of this mode is set by the adiabatic compressibility and probes the thermodynamic equation of state. In 2D, a dynamical scaling symmetry for atoms interacting via a $\delta$-potential predicts the breathing mode to occur at exactly twice the harmonic confinement frequency. However, a renormalized quantum treatment introduces a new length scale which breaks this classical scale invariance resulting in a so-called quantum anomaly. Our measurements deep in the 2D regime lie above the scale-invariant prediction for a range of interaction strengths indicating the breakdown of a $\delta$-potential model for atomic interactions. As the dimensionality is tuned from 2D to 3D we see the breathing oscillation frequency evolve smoothly towards the 3D limit. 

\end{abstract}
\pacs{03.75.Ss, 03.75.Hh, 05.30.Fk, 67.85.Lm}
\maketitle

Two-dimensional (2D) materials exhibit many novel physical properties \cite{castroneto09,fiori14,saito17,kou17}, but strong correlations and imperfections mean these are often difficult to understand theoretically. Quantum gases of neutral atoms may help address such fundamental challenges \cite{bloch08,levinsen15}, as well as new phenomena, not readily accessible in other systems. One scenario, generally encountered in quantum field theories, is anomalous symmetry breaking. Specifically, a \emph{quantum anomaly} occurs when a symmetry, present in a classical theory, is broken in the corresponding (renormalized) quantum theory. A paradigmatic example relevant to atomic collisions in ultracold alkali gases is the 2D $\delta$-potential \cite{mead91,hols93}, where an additional length scale associated with the interactions is required to remove divergences in the elementary theory. 

Anomalous symmetry breaking has been considered in the context of 2D harmonically confined Bose \cite{olshanii10} and Fermi gases \cite{hofmann12,taylor12,gao12} where interactions can be enhanced near a Feshbach resonance. In both cases, the anomaly leads to an increase in the frequency of the radial monopole mode, or breathing oscillation, above the value set by the scaling symmetry of the classical theory \cite{pitaevskii97}. Previous experiments have studied the radial breathing mode in 2D Bose \cite{merloti13a,merloti13b} and Fermi gases \cite{vogt12}, although the anomalous upshift has not yet been observed. More broadly, the breathing mode is a sensitive probe of the adiabatic compressibility and hence the thermodynamic equation of state \cite{dalfovo99,giorgini08} of the gas being studied.

In this Letter, we present measurements of the radial breathing mode frequency $\omega_B$ for highly oblate Fermi gases as a function of the interaction strength and dimensionality. The dimensionality of a harmonically trapped gas can be tuned by varying the chemical potential $\mu$ relative to the confinement energies $\hbar \omega_i$, ($i = x, y, z$) in each dimension. When $\hbar \omega_z \gg \mu, k_B T \gg \hbar \omega_{x,y}$, where $k_B$ is Boltzmann's constant and $T$ is the temperature, motion in the transverse ($z$) dimension can be frozen out and the gas becomes kinematically 2D. Pauli exclusion sets an upper limit on the total number of atoms $N < N_{2D}^{(Id.)}$ for an ideal (two-component) Fermi gas to remain 2D, where $N_{2D}^{(Id.)} \approx (\omega_z/ \omega_r)^2$ and $\omega_r = \sqrt{\omega_x \omega_y}$~\cite{dyke11}. Our measurements of $\omega_B$ in the deep 2D limit lie above the scale invariant prediction of $\omega_B = 2 \, \omega_r$ \cite{pitaevskii97} across a range of interaction strengths for $T \lesssim 0.2 \, T_F$, where $T_F $ is the Fermi temperature. As $N$ is increased, $\mu$ increases monotonically, reaching the 3D regime when $\mu \gg \hbar \omega_{x,y,z}$. The breathing mode frequency is seen to evolve smoothly through this crossover, from the 2D to the 3D limits. 

First, we consider gases prepared in the 2D limit. In ultracold collisions the range of the interatomic potential, $r_0$, is typically much smaller than the de Broglie wavelength, $\lambda_{dB}$. As such, $s$-wave collisions dominate and the detailed shape of the short-range potential has little impact on the (dilute) many-body system. One can then employ a simpler effective potential, typically a $\delta$-potential, that produces the correct scattering phase shift for $r \gg r_0$. Consider a 2D gas described by the hamiltonian,
\begin{equation} \label{eq:1} 
H(\mathbf{r}) = -\frac{\hbar^2}{2m}\nabla^2 - g \sum_{j<l} \delta^{(2)} (\mathbf{r}_j-\mathbf{r}_l),
\end{equation} 
where $m$ is the atomic mass, $g$ is a coupling constant and $\mathbf{r}_i$ is the position of the $i$-th atom. Crucially, this hamiltonian is scale invariant: this can be seen formally by replacing $\mathbf{r} \rightarrow \lambda \mathbf{r}$, which yields $H(\lambda \mathbf{r})\rightarrow H(\mathbf{r})/ \lambda^2$. Adding a harmonic potential, $V(\mathbf{r}) = \frac{1}{2} m\omega^2\mathbf{r}^2$, to (\ref{eq:1}) trivially breaks this scale invariance; however, it is replaced by a dynamical SO(2,1) scaling symmetry, as shown by Pitaevskii and Rosch \cite{pitaevskii97}, leading to a series of undamped excitations, with frequencies of precisely 2$j \omega_r$, where $j = 1,2...$, independent of $g$.

Quantum scattering in the 2D $\delta$-potential, however, is known to lead to divergences \cite{mead91,hols93}. Real atomic potentials support finite energy bound states and to correctly include this in a $\delta$-potential model one must renormalize the interaction thereby introducing a dimensionful parameter, $a_{2D}$, the (2D) $s$-wave scattering length. This interaction length scale explicitly breaks the classical scale invariance and the Pitaevskii-Rosch scaling symmetry, leading to an upshift in the breathing mode frequency \cite{olshanii10,hofmann12,taylor12,gao12}. In real experiments, the transverse confinement length, $\ell_z = \sqrt{\hbar/(m \omega_z)}$, introduces an additional length scale that must also be taken into account \cite{petrov01,merloti13a,levinsen15}. When the 3D scattering length $a_{3D}$ is much smaller in magnitude than $\ell_z$, the 2D coupling constant becomes independent of momentum $g \rightarrow \frac{\sqrt{8 \pi} \hbar^2 }{m} \frac{a_{3D}}{\ell_z}$ \cite{petrov00} and scale invariant behavior is recovered \cite{hung11,yefsah11,desbuqois14}. Using a Feshbach resonance to tune $a_{3D}$, one can produce strongly interacting gases where the departure from scale invariance will be more apparent. 

Scale invariant systems bear the property that the pressure, $P$, is proportional to the energy density ${\cal E}$. It is already evident, however, from the 2D Tan relation,
\begin{equation} \label{eq:2} 
P = {\cal E} +  \frac{\hbar^2}{4\pi m} {\cal C},
\end{equation} 
where ${\cal C} = \frac{\hbar^2}{2 \pi m} \frac{\partial{{\cal E}}}{\partial{\ln{(a_{2D})}}}$ is the contact density, that strict scale invariance should only exist when ${\cal C}$ vanishes (the noninteracting limit) \cite{tan08,valiente12,hofmann12,werner12}. Nonetheless, Taylor and Randeria showed that the anomalous frequency shift is smaller than might be anticipated from Eq.~(\ref{eq:2}) \cite{taylor12}. The energy of long wavelength modes (such as the breathing oscillation) is far too low to couple to pair-breaking excitations. As a result, the contribution to the contact related to the molecular bound state (which is always present for attractive interactions in 2D) does not affect the collective oscillation. Only the density dependent contribution to the contact affects to the compressibility, set by the derivative of (\ref{eq:2}) with respect to density $n \left ( \frac{\partial{P}}{\partial n} \right )_s$, at constant entropy density $s$. The largest deviation of $\omega_B$ from the scale invariant result is expected near the pole of the Feshbach resonance, where anomalous frequency shifts of up $10\%$ have been predicted at $T \rightarrow 0$ \cite{hofmann12,gao12}. Early investigations, however, were unable to resolve a shift at $T = 0.4 \, T_F$ \cite{vogt12} .

To produce single 2D $^6$Li Fermi gases, we load atoms into a hybrid optical and magnetic trap \cite{dyke16,fenech16} in a balanced mixture of the $|F = 1/2, m_F =  1/2\rangle$ and $|F = 3/2, m_F =  -3/2\rangle$ hyperfine states. A blue-detuned TEM$_{01}$ mode laser beam provides strong confinement along $z$ \cite{smith05} with $\omega_z/(2\pi) = 5.50 \pm 0.05$ kHz. Radial confinement arises from the residual curvature in the magnetic field produced by the Feshbach coils, leading to a highly harmonic potential with $\omega_r / (2 \pi) \approx 22 \,$Hz. The resultant aspect ratio, $\omega_z/\omega_r = 250$, gives $N_{2D}^{(Id.)} \approx 6 \times 10^4$. A broad Feshbach resonance at 690$\,$G is used to tune $a_{3D}$~\cite{zurn13}. In harmonically trapped 2D Fermi gases we parameterize the interactions by $\ln{(k_F^{\mathrm{HO}} a_{2D})}$, where $k_F^{\mathrm{HO}} = (4N)^{1/4}/\ell_r$ is the trap-averaged Fermi wavevector, $\ell_r = \sqrt{\hbar / (m \omega_r)}$ is the radial confinement length scale and $a_{2D}$ is defined as in~\cite{petrov01,makhalov14,boett16}.

\begin{figure}[!t]
	        \centering
	        \includegraphics[clip,width=0.47\textwidth]{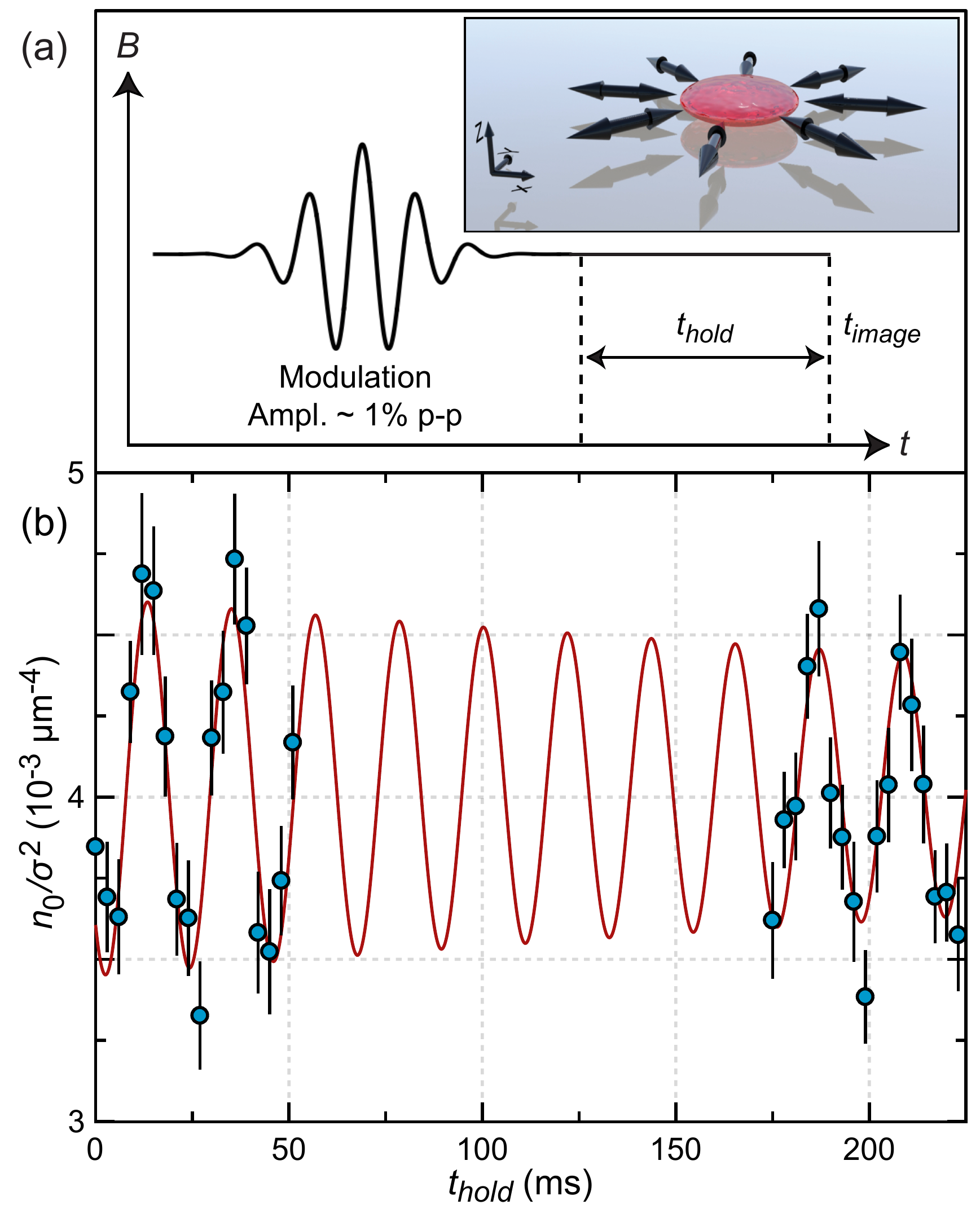}
	        \caption{(Color online) (a) The radial breathing mode is excited in the $x$-$y$ plane by applying a 5 cycle modulation with a triangular envelope to the magnetic field that provides the radial confinement. The modulation amplitude is chosen such that the cloud radius changes by a few percent after the driving. An \emph{in situ} absorption image of the cloud is taken after a variable hold time of up to 250$\,$ms. Inset: schematic of the breathing mode. (b) Representative data set showing $n_0/\sigma^2$ as a function of the hold time for $\ln(k_F^{\mathrm{HO}} a_{2D})$ = 3.3 with $N/N_{2D}^{(Id.)}$ = 0.3, as described in the text. This data yields $\omega_B/(2 \pi) = 46.03 \pm 0.35$ Hz. }
        \label{fig1}
\end{figure}

In a first set of experiments we study the breathing mode in 2D Fermi gases as a function of the interaction strength. Breathing oscillations are excited by modulating the magnetic field that provides the radial confinement for five cycles using a triangular envelope with a peak amplitude of $\sim 1\%$ of the total field, Fig.~\ref{fig1}(a). The modulation frequency is chosen to be close to the expected breathing frequency (although the measured $\omega_B$ is insensitive to $\pm 1 \,$Hz changes in the driving frequency). After the modulation, the cloud is held in the trap for a variable time, $t_{hold}$, before an absorption image is taken yielding the 2D density, $n(x,y)$. Due to the highly symmetric radial potential, an estimate of the cloud width and peak density can be found by fitting a Gaussian to the azimuthal average of $n(x,y)$. To optimize signal to noise, we plot the ratio of the peak density, $n_0$, to the square of the cloud radius, $\sigma^2$ (which to first order is insensitive to atom number fluctuations), as a function of $t_{hold}$. The maximum change in peak density $\delta n_0/n_0$ is typically $ \sim 6\%$ and in cloud width $\delta\sigma /\sigma \sim 3\%$. We obtain three to five sets of $n_0 / \sigma^2$ vs.~$t_{hold}$ data, ensuring the atom number varies by less than $10\%$ across all measurements. We then fit a damped sinusoidal function to the average of these to determine $\omega_B$. Between each $n_0 / \sigma^2$ vs.~$t_{hold}$ data set, we also measure the in-plane trapping frequencies, $\omega_x$ and $\omega_y$, parallel and perpendicular to the propagation axis of the TEM$_{01}$ mode laser, by exciting center of mass oscillations. This allows for accurate determination of $\omega_r$ which is necessary for subsequent analysis \cite{drift}. A representative collective oscillation measurement and fit for $\ln(k_F^{\mathrm{HO}} a_{2D})$ = 3.3 is shown in Fig.~\ref{fig1}(b). Error bars indicate the standard deviation of $n_0 / \sigma^2$ at each $t_{hold}$. 

\begin{figure}[!t]
        \centering
        \includegraphics[clip,width=0.47\textwidth]{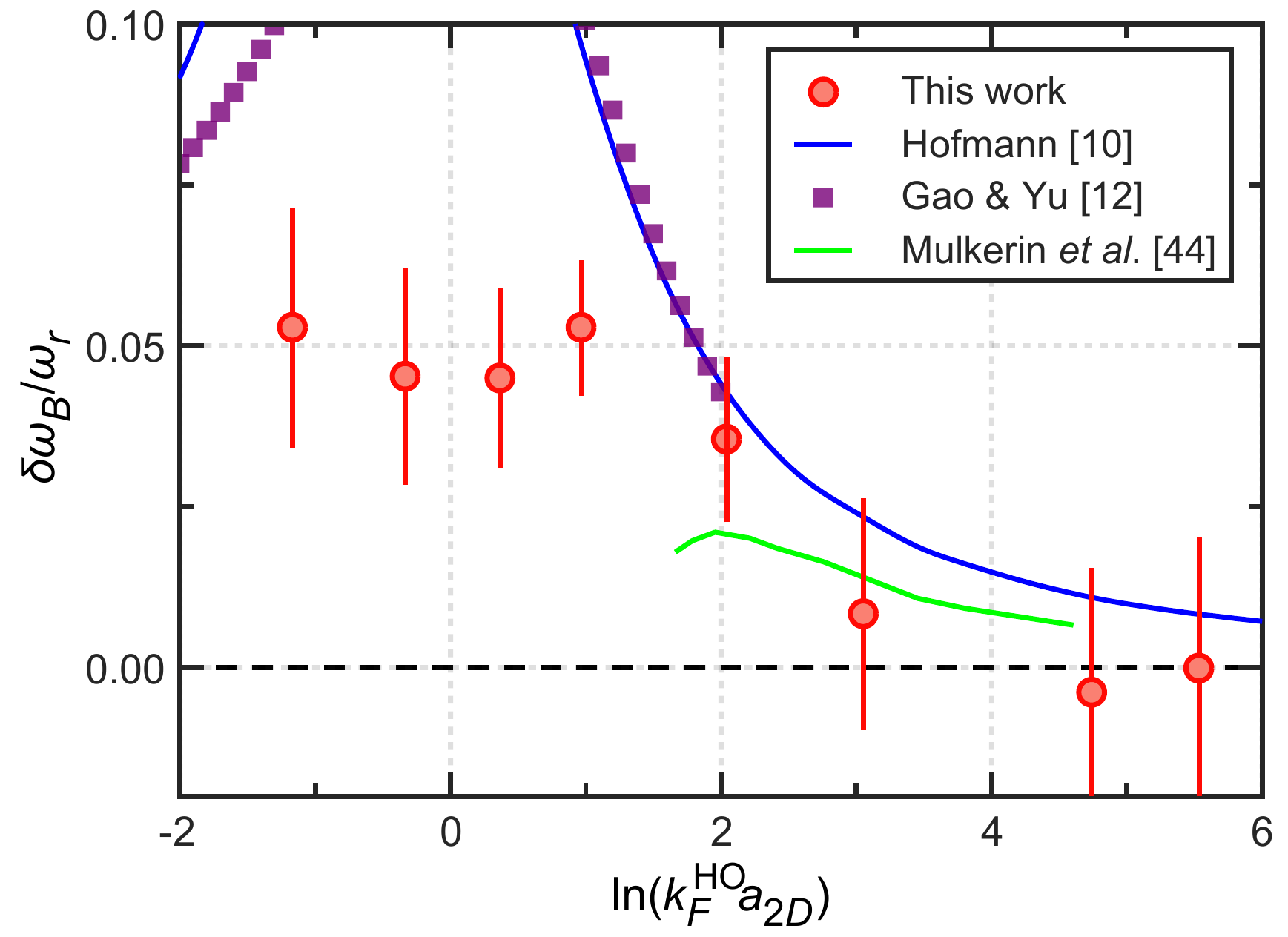}
        \caption{(Color online) Relative shift of the breathing mode frequency $\delta \omega_B/\omega_r \equiv (\omega_B - 2 \, \omega_r)/\omega_r$ from the scale invariant prediction of $2 \, \omega_r$ (black dashed line) as a function of the interaction strength $\ln(k_F^{\mathrm{HO}} a_{2D})$.  The temperatures of all clouds used for the above data lie in the range of $T/T_F = 0.14 - 0.22$.}
        \label{fig2}
\end{figure}

In Fig.~\ref{fig2} we plot the relative deviation of the breathing mode frequency, $\delta \omega_B / \omega_r = \omega_B / \omega_r - 2$, for $N \approx 16 \times 10^3 \, (\leq 0.3 N_{2D}^{(Id.)})$ as a function of $\ln{(k_F^{\mathrm{HO}} a_{2D})}$. In the BCS limit, when $\ln(k_F^{\mathrm{HO}} a_{2D})$ is large and positive, the breathing mode frequency is consistent with the scale invariant prediction 2$\, \omega_r$. In the weakly interacting regime, $\ln{(k_F^{\mathrm{HO}} a_{2D})} \gtrsim 5$, the contact is small and the anomalous shift should be less than 1$\%$ \cite{hofmann12,gao12}. Closer to the Feshbach resonance, for $\ln{(k_F^{\mathrm{HO}} a_{2D})} \lesssim 3$, we observe $\omega_B$ to increase, reaching approximately 2.05$\, \omega_r$ in the strongly interacting regime, $|  \ln{(k_F^{\mathrm{HO}} a_{2D})}  | \lesssim 1$. This represents a clear signature of the anomalous upshift. 

Cloud temperatures for the data in Fig.~\ref{fig2} were in the range $0.14 \leq T/T_F \leq 0.22$, where $k_B T_F = E_F = \frac{\hbar^2 \pi}{m} n_0$ is the Fermi energy at the trap center. Our thermometry is based on fitting $n(x,y)$ to the equation of state, $n(\mathbf{r}) \lambda_{dB}^2 = f_n[\beta \mu(\mathbf{r}), \beta E_b]$, where $\lambda_{dB} = \sqrt{2 \pi \hbar^2 \beta / m}$, $\beta = 1/(k_B T)$, $\mu(\mathbf{r}) = \mu(0) - V(\mathbf{r})$, is the chemical potential in the local density approximation (LDA), $E_b = \hbar^2/(m a_{2D}^2)$ is the molecular binding energy and $f_n$ is a universal function given by the virial expansion for $\beta \mu \leq -3.5$ \cite{liu10} and a self-consistent (GG) $t$-matrix calculation for $-3.5 < \beta \mu < 0.5$ \cite{bauer14,mulkerin15}. As temperature enters $f_n[\beta \mu(\mathbf{r}), \beta E_b]$ via both the scaled interaction energy and chemical potential, we use an iterative fitting routine based on a bisection algorithm, that fits $\mu(0)$ and $T$ for different $\beta E_b$, until $\beta E_b$ converges with the fitted $T$. While the GG calculation is not exact, comparison with experiments \cite{mulkerin15,fenech16} indicate that the size of any systematics should be small compared to the $\sim 15 \%$ uncertainty in the fitted temperatures. 

Having identified the anomalous shift in the 2D limit, we now consider the evolution of the breathing mode frequency through the 2D-3D crossover as the atom number is increased. Again, $\omega_B/\omega_r$ serves as a sensitive \emph{in situ} probe of the relationship between pressure and density. Consider gases described a polytropic equation of state,
\begin{equation} \label{eq:5} 
P(s) = n^q p(s),
\end{equation} 
where $q$ is the polytropic coefficient (which can depend on the dimensionality, interactions, temperature and quantum statistics), and $p(s)$ is a ($T$-dependent) function of the entropy. Equations of state of this form permit simple solutions to the hydrodynamic equations, valid for both superfluids and normal phase gases in the collisional regime. For the 2D breathing mode one obtains $\omega_B = \sqrt{2 q} \, \omega_r$ \cite{heiselberg04,derosi15}. Of relevance here are the strict 2D limit, where $\hbar \omega_z \gg \mu$ and $q \rightarrow 2$ (ignoring the small anomalous upshift), and the oblate 3D limit, where $\mu \gg \hbar \omega_{z} \gg \hbar \omega_B$. The latter case refers to gases that are thermodynamically 3D, where the wavelength of the collective oscillation (on the order of the radial cloud size, $\ell_{r}^2 k_F^{\mathrm{HO}}$) is much larger than the transverse cloud size ($\ell_{z}^2 k_F^{\mathrm{HO}}$), such that the breathing excitation is dynamically two-dimensional. In this limit, one can use the LDA, $\mu(x,y,z) = \mu(x,y,0) - V(z)$, to integrate over $z$ and recover an effective 2D equation of state for the collective oscillation. For a Fermi gas in the unitarity limit ($a_{3D} \rightarrow \infty$) one obtains $q = 3/2$ and $\omega_B = \sqrt{3} \, \omega_r$ \cite{hu14,derosi15}. Thus, the crossover from a thermodynamically 2D to 3D Fermi gas with resonant interactions should be marked by a change in $\omega_B / \omega_r$ from $\sim 2$ to $\sqrt{3}$ \cite{toniolo18}. 

\begin{figure}[!t]
	\centering
		\includegraphics[clip, width=0.46\textwidth]{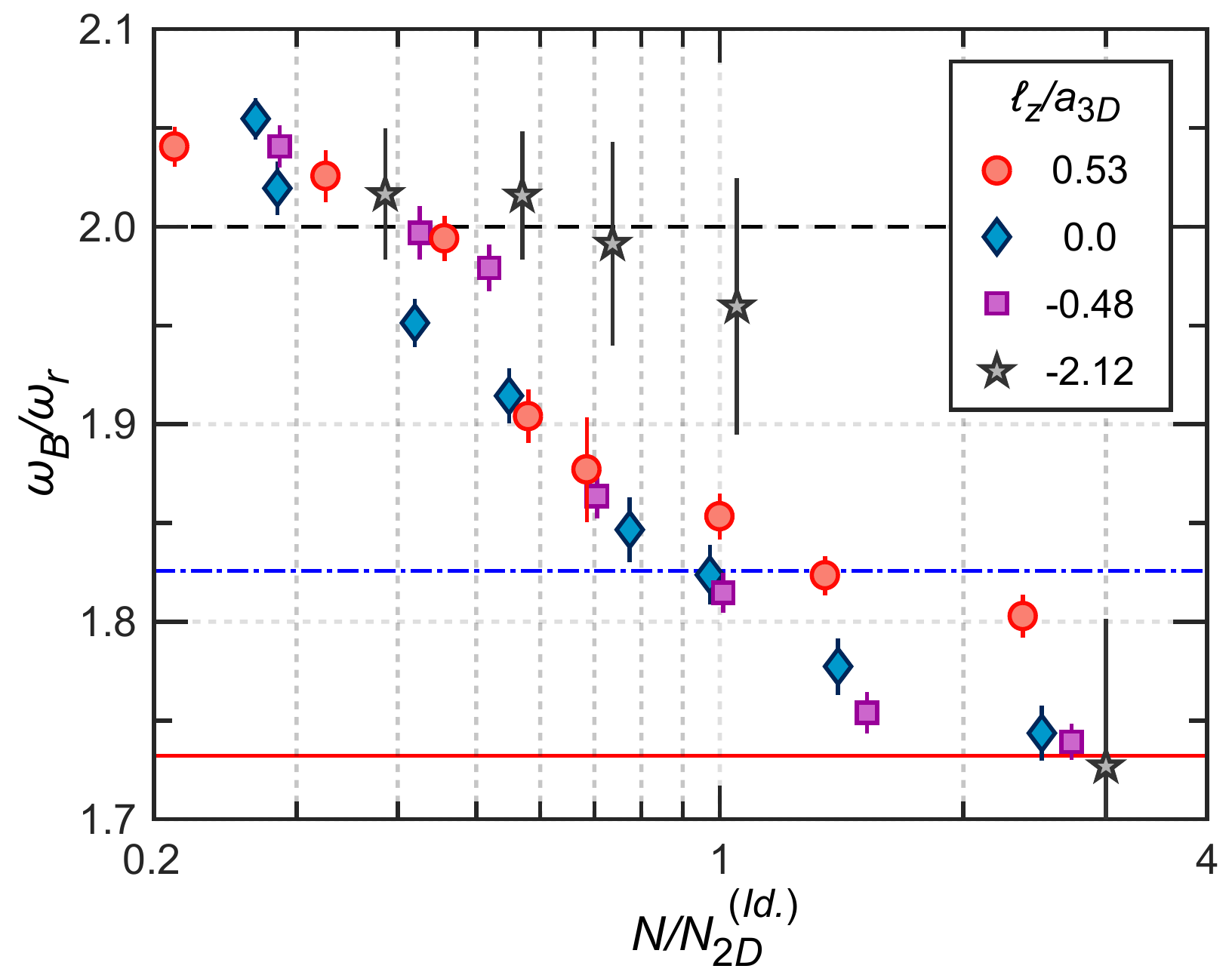}
		\caption{(Color online) Frequency of the breathing mode through the 2D to 3D crossover for, $\ell_z/a_{3D} = 0.53$ (red circles), $0$ (blue diamonds), $-0.48$ (purple squares) and $-2.12$ (grey stars). The black dashed line indicates the scale invariant 2D breathing mode frequency $2 \, \omega_r$, red solid line indicates the theoretical breathing mode frequency $\sqrt{3} \, \omega_r$ for a unitary Fermi gas in the 3D thermodynamic limit and the blue dot-dashed line $\sqrt{10/3} \, \omega_r$ shows the limit for a weakly interacting (thermodynamically 3D) Bose gas.}
		\label{fig3}
\end{figure}

To investigate this, we study the breathing oscillation as a function of $N$ for various interaction strengths, using the procedure described above (Fig.~\ref{fig1}). Figure \ref{fig3} displays $\omega_B/\omega_r$ for four different values of $\ell_z/a_{3D}$, as a function of $N/N_{2D}^{(Id.)}$. The three strongly interacting clouds ($| \ell_z / a_{3D} | < 1$) all show similar behavior. Deep in the 2D regime, $\omega_B$ lies above 2$\, \omega_r$ consistent with the anomalous upshift, however, as $N$ increases, $\omega_B/\omega_r$ steadily shifts below 2, reaching the 3D limit at $N/N_{2D}^{(Id.)} \approx 3$. In contrast, for the cloud with weaker interactions, $\ell_z / a_{3D} = -2.12$, $\omega_B/\omega_r$ remains close to 2 for all $N/N_{2D}^{(Id.)} \lesssim 1$, and only decreases significantly at the highest atom number. All of the large $N$ clouds have high peak densities and thus approach the strongly interacting regime in the 3D limit where behavior similar to a unitary gas can be expected. The largest cloud for $\ell_z / a_{3D} \approx +0.5$ shows a possible deviation towards the bosonic molecule result $\omega_B/\omega_r = \sqrt{10/3}$ \cite{derosi15}.

An interesting feature evident in Fig.~\ref{fig3} is that $\omega_B$ begins falling immediately with increasing $N$ in the strongly interacting clouds. This suggests these clouds may not strictly satisfy the 2D limit, even though $N < N_{2D}^{(Id.)}$. As seen in previous studies, finite transverse confinement can influence the dynamics of quasi-2D gases while $\mu < \hbar \omega_z$ \cite{merloti13a,dyke16}. Many-body effects may enhance pairing in the normal phase \cite{murthy18} meaning the condition $E_b \ll \hbar \omega_z$ is less easily satisfied. On the other hand, 2D models suggest that finite temperatures can also drive the breathing mode frequency below 2 close to a Feshbach resonance \cite{chafin13}, even for $T \approx 0.2 \, T_F$ \cite{mulkerin17}. Both nonzero temperatures and finite transverse confinement may be limiting the observed increase above $2 \, \omega_r$ in the strongly interacting clouds. 

In summary, we have studied the monopole breathing mode frequencies of an interacting 2D Fermi gas throughout the BEC-BCS crossover. In the deep 2D regime we observe a departure above the scale invariant providing evidence of the quantum anomaly. This measurement indicates a breakdown of $\delta$-potential models in alkali atomic gases. We observe a shift in $\omega_B$ of $\sim 2.5\%$ above the scale-invariant value for the strongest interactions. We have also measured the breathing mode frequency through the 2D to 3D crossover at various interaction strengths and seen how the gas evolves from a 2D to 3D thermodynamic equation of state. Understanding the role of finite temperatures and the transverse dimension in strongly interacting gases represents a future challenge in these systems \cite{toniolo17}.

We note that results similar to those presented here have recently been reported \cite{holten18}. 

We would like to thank P. Hannaford, M. Parish, J. Levinsen and R. Fletcher for helpful discussions. C. J. V. and P. D. acknowledge financial support from the Australian Research Council Programs No. FT120100034, No. DE140100647 and CE170100039.


\begin{thebibliography}{10}
\bibitem{drift} Day-to-day drifts in the alignment of the TEM$_{01}$ mode laser beam can lead to measured asymmetries in the radial potential of up to $2\%$.
\end{thebibliography}
\end{document}